\documentclass[10pt,conference]{IEEEtran}

\usepackage{bm}
\usepackage{amsmath}
\usepackage{amssymb}
\newfont{\boldit}{cmbxti10}

\newtheorem{theorem}{Theorem}
\newtheorem{lemma}{Lemma}

\begin{document}

\title{Uncorrectable Errors of Weight Half the Minimum Distance for Binary Linear Codes}



\author{
\authorblockN{Kenji Yasunaga}
\authorblockA{Graduate School of Science and Technology\\
Kwansei Gakuin University\\
2-1 Gakuen, Sanda, 669-1337 Japan\\
E-mail: yasunaga@kwansei.ac.jp}
\and
\authorblockN{Toru Fujiwara}
\authorblockA{Graduate School of Information Science and Technology\\
Osaka University\\
1-5 Yamadaoka, Suita, 565-0871 Japan\\
E-mail: fujiwara@ist.osaka-u.ac.jp}
}

\maketitle

\begin{abstract}
A lower bound on the number of uncorrectable errors of weight half the minimum
distance is derived for binary linear codes satisfying some condition.
The condition is satisfied by some primitive BCH codes, extended primitive BCH
codes, Reed-Muller codes, and random linear codes.
The bound asymptotically coincides with the corresponding upper bound for
Reed-Muller codes and random linear codes.
By generalizing the idea of the lower bound,
a lower bound on the number of uncorrectable errors for weights larger than
half the minimum distance is also obtained, but the generalized lower bound is
weak for large weights.
The monotone error structure and its related notion {\em larger half\/} and {\em trial set\/}, 
which are introduced by Helleseth, Kl\o ve, and Levenshtein, are mainly used to derive the bounds.
\end{abstract}

\section{Introduction}
For a binary linear code, correctable errors we consider here are binary errors correctable by the minimum distance decoding,
which performs a maximum likelihood decoding for binary symmetric channels.
Syndrome decoding is one of the minimum distance decoding.
In syndrome decoding, the correctable errors are coset leaders of the code.
When there are two or more minimum weight vectors in a coset, 
we have choices of the coset leader.
If the lexicographically smallest minimum weight vector is taken as the coset leader,
then both the correctable errors and the uncorrectable errors have a monotone structure.
That is, when $\bm{y}$ covers $\bm{x}$ (the support of $\bm{y}$ contains that of $\bm{x}$),
if $\bm{y}$ is correctable, then $\bm{x}$ is also correctable, and 
if $\bm{x}$ is uncorrectable, then $\bm{y}$ is also uncorrectable~{\cite{peterson}}. 
Using this monotone structure, Z\'emor showed that the residual error probability after maximum likelihood decoding 
displays a threshold behavior~\cite{zemor93}.
When uncorrectable (and correctable) errors have the monotone structure, they are characterized by the minimal uncorrectable
(and maximal correctable) errors.
Larger halves of codewords are introduced by Helleseth, Kl\o ve, and Levenshtein~\cite{helleseth05} 
to describe the minimal uncorrectable errors.
They also introduced a trial set for a code. It is a set of codewords whose larger halves contain all minimal uncorrectable errors.
Trial sets can be used for a maximum likelihood decoding and for giving an upper bound on the number of uncorrectable errors.
The set of all codewords except for the all-zero codeword and the set of minimal codewords~\cite{ashikhmin98} in the code 
are examples of trial sets.

In this paper, we study bounds on the number of correctable/uncorrectable
errors.
There were several works about them.
For the first-order Reed-Muller codes, the exact numbers of correctable errors of weight half the minimum distance and
half the minimum distance plus one were determined~\cite{wu98, yasunaga07}.
For general linear codes, some upper bounds on the number of uncorrectable errors were presented 
in~\cite{helleseth05, helleseth97, poltyrev94}.
In this work, we consider lower bounds on the number of uncorrectable errors based on the idea of~\cite{helleseth05}
for general linear codes.

We derive a lower bound on the number of uncorrectable errors of weight half the minimum distance for codes satisfying some condition.
The bound is given in terms of the numbers of codewords with weights $d$ and $d+1$ in a trial set for odd $d$,
where $d$ is the minimum distance of the code.
For the case of even $d$, the bound is given by the number of codewords with weight $d$ in a trial set.
Since the set of all codewords except the all-zero vector is a trial set,
the bound can be evaluated by the numbers of codewords with weights $d$ and $d+1$.
The condition is not too restrictive, and some primitive BCH codes, extended primitive BCH codes, Reed-Muller codes,
and random linear codes satisfy the condition.
For Reed-Muller codes and random linear codes, the lower bound asymptotically coincides with the upper bound of~\cite[Corollary~7]{helleseth05}.
The lower bound can be generalized to a lower bound on the size of the set of larger halves of a trial set, 
which is a lower bound on the number of uncorrectable errors.

In the next section, we review some definitions and properties of the monotone error structure, larger halves, and trial sets.
In Section~\ref{sec:halfmd}, a lower bound on the number of uncorrectable errors of weight half the minimum distance is
given for the codes satisfying some condition.
The bound presented in Section~\ref{sec:halfmd} is generalized in Section~\ref{sec:generalize}.

\section{Larger halves and trial sets}

We introduce definitions and properties of larger halves and trial sets.
Let $\mathbb{F}^n$ be the set of all binary vectors of length $n$.
Let $C \subseteq \mathbb{F}^n$ be a binary linear code of length $n$, dimension $k$, and minimum distance $d$.
Then $\mathbb{F}^n$ is partitioned into $2^{n-k}$ cosets $C_1, C_2, \ldots, C_{2^{n-k}}$;
$\mathbb{F}^n = \bigcup_{i=1}^{2^{n-k}}C_i$ and $C_i \cap C_j = \emptyset$ for $i \neq j$,
where each $C_i=\{\bm{v}_i+\bm{c} : \bm{c} \in C\}$ with $\bm{v}_i \in \mathbb{F}^n$.
The vector $\bm{v}_i$ is called a coset leader of the coset $C_i$ if the weight of $\bm{v}_i$ is smallest in $C_i$.

Let $H$ be a parity check matrix of $C$.
The syndrome of a vector $\bm{v} \in \mathbb{F}^n$ is defined as $\bm{v}H^T$.
All vectors having the same syndrome are in the same coset.
Syndrome decoding associates an error vector to each syndrome.
The syndrome decoder presumes that the error vector added to the received vector $\bm{y}$
is the coset leader of the coset which contains $\bm{y}$.
The syndrome decoding function $D : \mathbb{F}^n \rightarrow C$ is defined as
\begin{equation*}
D(\bm{y}) = \bm{y} + \bm{v}_i, \ \ \text{if} \ \bm{y} \in C_i.
\end{equation*}

In this paper, we take as $\bm{v}_i$ the minimum element in $C_i$ with respect to the following total ordering $\preceq$:
\begin{equation}
\bm{x} \preceq \bm{y} \ \ \text{if and only if} \ \
\left\{ 
\begin{array}{l}
w(\bm{x}) < w(\bm{y}), \ \ \text{or} \\
w(\bm{x}) = w(\bm{y}) \ \text{and} \ v(\bm{x}) \leq v(\bm{y}),
\end{array}
\right.\nonumber
\end{equation}
where $w(\bm{x})$ denotes the Hamming weight of a vector $\bm{x}=(x_1, x_2, \ldots, x_n)$ and
$v(\bm{x})$ denotes the numerical value of $\bm{x}$: 
\begin{equation*}
v(\bm{x}) = \sum_{i=1}^{n} x_i 2^{n-i}.
\end{equation*}
We write $\bm{x} \prec \bm{y}$ if $\bm{x} \preceq \bm{y}$ and $\bm{x} \neq \bm{y}$.

Let $E^0(C)$ be the set of all coset leaders of $C$.
In the syndrome decoding, $E^0(C)$ is the set of correctable errors and
$E^1(C) = \mathbb{F}^n \setminus E^0(C)$ is the set of uncorrectable errors.
Since we take the minimum element with respect to $\preceq$ in each coset as its coset leader,
both $E^0(C)$ and $E^1(C)$ have the following well-known monotone structure (see~{\cite[Theorem 3.11]{peterson}}).
Let $\subseteq$ denote a partial ordering called ``covering'' such that
\begin{equation}
\bm{x} \subseteq \bm{y} \ \text{ if and only if} \  S(\bm{x}) \subseteq S(\bm{y}),\nonumber
\end{equation}
where
\begin{equation*}
S(\bm{v})=\{ i : v_i \neq 0\}
\end{equation*}
is the support of $\bm{v}=(v_1,v_2,\ldots,v_n)$.
Consider $\bm{x}$ and $\bm{y}$ with $\bm{x} \subseteq \bm{y}$. 
If $\bm{y}$ is a correctable error, then $\bm{x}$ is also correctable.
If $\bm{x}$ is uncorrectable, then $\bm{y}$ is also uncorrectable.
Using this structure, Z\'emor showed that the residual error probability after maximum likelihood decoding
displays a threshold behavior~\cite{zemor93}.
Helleseth, Kl\o ve, and Levenshtein~\cite{helleseth05} studied this structure and introduced {\em larger halves\/} and {\em trial sets\/}.

Since the set of uncorrectable errors $E^1(C)$ has a monotone structure,
$E^1(C)$ can be characterized by {\em minimal uncorrectable errors\/} in $E^1(C)$.
An uncorrectable error $\bm{y} \in E^1(C)$ is minimal if there exists no $\bm{x}$ such that
$\bm{x} \subset \bm{y}$ in $E^1(C)$.
We denote by $M^1(C)$ the set of all minimal uncorrectable errors in $C$.
Larger halves of a codeword $\bm{c} \in C$ are introduced to characterize the minimal uncorrectable errors,
and are defined as minimal vectors $\bm{v}$ with respect to covering such that $\bm{v} + \bm{c} \prec \bm{v}$.
The following condition is a necessary and sufficient condition that $\bm{v} \in \mathbb{F}^n$ 
is a larger half of $\bm{c} \in C$:
\begin{gather}
\bm{v}  \subseteq \bm{c},\label{eq:lhcond1}\\
w(\bm{c})  \leq 2w(\bm{v}) \leq w(\bm{c})+2,\label{eq:lhcond2}\\
l(\bm{v})  \begin{cases}
= l(\bm{c}) & \text{if} \ \ 2w(\bm{v}) = w(\bm{c}),\\
> l(\bm{c}) & \text{if} \ \ 2w(\bm{v}) = w(\bm{c})+2,
\end{cases}\label{eq:lhcond3}
\end{gather}
where
\begin{equation}
l(\bm{x}) = \min S(\bm{x}),
\end{equation}
that is, $l(\bm{x})$ is the leftmost non-zero coordinate in the vector $\bm{x}$.
The condition~(\ref{eq:lhcond3}) is not applied if $w(\bm{c})$ is odd.
The proof of equivalence between the definition and the above condition is found in the proof of
\cite[Theorem~1]{helleseth05}.
Let $LH(\bm{c})$ be the set of all larger halves of $\bm{c} \in C$.
For a set $U \subseteq C \setminus \{ \bm{0} \}$, define 
\begin{equation*}
LH(U) = \bigcup_{\bm{c} \in U} LH(\bm{c}).
\end{equation*}
When the weight of a codeword $\bm{c}$ is odd, the weight of the vectors in $LH(\bm{c})$ is $(w(\bm{c})+1)/2$.
When the weight of $\bm{c}$ is even, $LH(\bm{c})$ consists of vectors of weights $w(\bm{c})/2$ and $w(\bm{c})/2+1$.
For convenience, let $LH^-(\bm{c})$ and $LH^+(\bm{c})$ denote the sets of larger halves of $\bm{c}$ of weight 
$w(\bm{c})/2$ and $w(\bm{c})/2+1$, respectively.
Then $LH(\bm{c}) = LH^-(\bm{c}) \cup LH^+(\bm{c})$.
Also let $LH^-(U) = \bigcup_{\bm{c} \in U}LH^-(\bm{c})$ and $LH^+(U) = \bigcup_{ \bm{c} \in U} LH^+(\bm{c})$ 
for a subset $U$ of even-weight subcode.

A trial set $T$ for a code $C$ is defined as follows:
\begin{equation*}
T \subseteq C \setminus \{\bm{0}\} \text{ is a trial set for } C \ \text{if} \ M^1(C) \subseteq LH(T).
\end{equation*}
Since every larger half is an uncorrectable error, we have the relation
\begin{equation}
M^1(C) \subseteq LH(T) \subseteq E^1(C).\label{eq:mlhe}
\end{equation}

In the rest of paper, for $\bm{u}, \bm{v} \in \mathbb{F}^n$, we write $\bm{u} \cap \bm{v}$ as the vector in $\mathbb{F}^n$
whose support is ${\rm S}(\bm{u}) \cap{\rm S}(\bm{v})$.
For a set $U \subseteq \mathbb{F}^n$, define
\begin{equation*}
A_i(U)=\{ \bm{v} \in U : w(\bm{v})=i \}.
\end{equation*}
Also we define $M^1_i(C) = A_i(M^1(C))$ and $LH_i(U) = A_i(LH(U))$ for $U \subseteq C \setminus \{ \bm{0} \}$.

\section{A Bound on the Number of Uncorrectable Errors of Weight Half the Minimum Distance}\label{sec:halfmd}

In this section, we derive a lower bound on $|E^1_{\lceil d/2 \rceil}(C)|$.
The bound is given by the number of codewords with weights $d$ and $d+1$ in a trial set.
Since $C \setminus \{ \bm{0} \}$ is a trial set for $C$, the lower bound can be evaluated by the number of codewords of weights $d$ and $d+1$ in $C$.

Since the weight $\lceil d/2 \rceil$ is the minimum weight in $E^1(C)$, every vector in $E^1_{\lceil d/2 \rceil}(C)$ is not covered
by other uncorrectable errors, and thus $M^1_{\lceil d/2 \rceil}(C) = E^1_{\lceil d/2 \rceil}(C)$.
From~(\ref{eq:mlhe}), we have
\begin{equation*}
M^1_{\lceil d/2 \rceil}(C) = LH_{\lceil d/2 \rceil}(T) = E^1_{\lceil d/2 \rceil}(C),
\end{equation*}
where $T$ is a trial set for $C$.
We will give a lower bound on $|E^1_{\lceil d/2 \rceil}(C)|$ by giving a lower bound on $|LH_{\lceil d/2 \rceil}(T)|$.

\subsection{Odd Minimum Weight Case}

When $d$ is odd, $LH_{\lceil d/2 \rceil}(T) = LH(A_d(T)) \cup LH^-(A_{d+1}(T))$.
The next lemma implies that the number of common larger halves among $LH(A_d(T))$ and $LH^-(A_{d+1}(T))$ is small.
\medskip
\begin{lemma}\label{lem:oddcommon}
Let $C$ be a linear code with odd minimum distance $d$.
For every $\bm{c}_1, \bm{c}_1' \in A_d(C)$ and $\bm{c}_2, \bm{c}_2' \in A_{d+1}(C)$, it holds that
$|LH(\bm{c}_1) \cap LH(\bm{c}_1')| = 0$, $|LH(\bm{c}_1) \cap LH^-(\bm{c}_2)| \leq 1$, and $|LH^-(\bm{c}_2) \cap LH^-(\bm{c}_2')| \leq 1$.
\end{lemma}
\medskip
\begin{proof}
For $\bm{c}, \bm{c}' \in C \setminus \{ \bm{0} \}$,
every vector $\bm{v} \in LH(\bm{c}) \cap LH(\bm{c}')$ has the property that $\bm{v} \subseteq \bm{c} \cap \bm{c}'$.
Since every vector in $LH(\bm{c}_1), LH(\bm{c}_1'), LH^-(\bm{c}_2), LH^-(\bm{c}_2')$ has weight $(d+1)/2$,
it is enough to show that $w(\bm{c}_1 \cap \bm{c}_1') < (d+1)/2$, $w(\bm{c}_1 \cap \bm{c}_2) \leq (d+1)/2$, 
$w(\bm{c}_2 \cap \bm{c}_2') \leq (d+1)/2$.
We can prove them by using $w(\bm{c} \cap \bm{c}') = (w(\bm{c})+w(\bm{c}')-w(\bm{c}+\bm{c}'))/2$ and $w(\bm{c}+\bm{c}') \geq d$.



\end{proof}
\medskip

From the previous lemma, we give a lower bound on the number of uncorrectable errors of weight half the minimum distance.
The corresponding upper bound is given unconditionally by~\cite[Corollary~7]{helleseth05}.
\medskip
\begin{theorem}\label{th:oddlower}
Let $C$ be a linear code with odd minimum distance $d$ and $T$ be a trial set for $C$.
If 
\begin{equation}
\binom{d}{\frac{d+1}{2}} > |A_d(T)|+|A_{d+1}(T)|-1\label{eq:oddcond}
\end{equation} 
holds, then
\begin{multline*}
\binom{d}{\frac{d+1}{2}}(|A_d(T)|+|A_{d+1}(T)|)\\ - (2|A_d(T)|+|A_{d+1}(T)|-1)|A_{d+1}(T)|\\
\leq |E^1_{\frac{d+1}{2}}(C)|
\leq \binom{d}{\frac{d+1}{2}}(|A_d(T)|+|A_{d+1}(T)|).
\end{multline*} 
\end{theorem}
\medskip
\begin{proof}
From Lemma~\ref{lem:oddcommon}, a codeword $\bm{c} \in A_d(T)$ has at most one common larger half for every $\bm{c}' \in A_{d+1}(T)$
and does not have common larger halves for any $\bm{c}' \in A_{d}(T) \setminus \{ \bm{c} \}$.
Thus at least $|LH(\bm{c})| - |A_{d+1}(T)|$ vectors in $LH(\bm{c})$ does not have common larger halves.
Also, a codeword $\bm{c} \in A_{d+1}(T)$ has at most one common larger half for every $\bm{c}' \in A_{d}(T) \cup \{ A_{d+1}(T) \setminus \{\bm{c}\}\}$,
at least $|LH^-(\bm{c})| - |A_d(T)| - |A_{d+1}(T)| +1$ vectors in $LH^-(\bm{c})$ does not have common larger halves.

For every $\bm{c}_1 \in A_d(T)$ and $\bm{c}_2 \in A_{d+1}(T)$, we have $|LH(\bm{c}_1)|=|LH^-(\bm{c}_2)|=\binom{d}{(d+1)/2}.$
Therefore we have the lower bound $(\binom{d}{(d+1)/2} - |A_{d+1}(T)|)|A_d(T)| 
+ (\binom{d}{(d+1)/2} - |A_d(T)| - |A_{d+1}(T)| + 1)|A_{d+1}(T)| \leq |LH_{(d+1)/2}(T)| = |E^1_{(d+1)/2}(C)|$.
The upper bound is obtained from the inequality $|LH_{(d+1)/2}(T)| = |LH(A_d(T)) \cup LH^-(A_{d+1}(T))| \leq |LH(A_d(T))| + |LH^-(A_{d+1}(T))|
\leq \binom{d}{(d+1)/2}|A_d(T)| + \binom{d}{(d+1)/2}|A_{d+1}(T)|$.
\end{proof}
\medskip
The difference between the upper and lower bounds is $(2|A_d(C)|+|A_{d+1}(C)|-1)|A_{d+1}(C)|$.
If the fraction $|A_{d+1}(C)|/\binom{d}{(d+1)/2}$ tends to zero as the code length becomes large,
the lower bound asymptotically coincides with the upper one.

\subsection{Even Minimum Weight Case}

When $d$ is even, $LH_{\lceil d/2 \rceil}(T) = LH^-(A_d(T))$.
The next lemma implies that the number of common larger halves among $LH^-(A_d(T))$ is small.
\medskip
\begin{lemma}\label{lem:evencommon}
Let $C$ be a linear code with even minimum distance $d$.
For every $\bm{c}_1, \bm{c}_2 \in A_d(C)$, it holds that $|LH^-(\bm{c}_1) \cap LH^-(\bm{c}_2)| \leq 1$.
\end{lemma}
\medskip
\begin{proof}
For contradiction, suppose that there exist two distinct vectors in $LH^-(\bm{c}_1) \cap LH^-(\bm{c}_2)$.
Then it holds that $w(\bm{c}_1 \cap \bm{c}_2) \geq d/2+1$, but this leads to the contradiction that
 $w(\bm{c}_1 + \bm{c}_2) = w(\bm{c}_1) + w(\bm{c}_2) - 2w(\bm{c}_1 \cap \bm{c}_2) \leq d-1$.
\end{proof}
\medskip
\begin{theorem}\label{th:evenlower}
Let $C$ be a linear code with even minimum distance $d$.
If 
\begin{equation}
\frac{1}{2}\binom{d}{\frac{d}{2}} > |A_d(T)|-1\label{eq:evencond}
\end{equation} 
holds, then
\begin{multline*}
\frac{1}{2}\binom{d}{\frac{d}{2}}|A_d(T)| - (|A_d(T)|-1) |A_{d}(T)|\\
 \leq |E^1_{\frac{d}{2}}(C)|
\leq \frac{1}{2}\binom{d}{\frac{d}{2}}|A_d(T)|.
\end{multline*} 
\end{theorem}
\medskip
\begin{proof}
From Lemma~\ref{lem:evencommon}, a codeword $\bm{c} \in A_d(T)$ has at most one common larger half for every 
$\bm{c}' \in A_{d}(T) \setminus \{ \bm{c} \}$.
Thus at least $|LH^-(\bm{c})| - |A_{d}(T)| + 1$ vectors in $LH^-(\bm{c})$ does not have common larger halves.
Thus we have the lower bound $(\binom{d}{d/2}/2-|A_d(T)|+1)|A_d(T)| \leq |LH^-(A_d(T))| = |E^1_{d/2}(C)|$.
The upper bound is obtained from the inequality $|E^1_{d/2}(C)| = |LH^-(A_d(C))| \leq \binom{d}{d/2}|A_d(C)|/2$.
\end{proof}
\medskip

The difference between the upper and lower bounds is upper bounded by $|A_{d}(C)|^2$.
If the fraction $|A_{d}(C)|/\binom{d}{d/2}$ tends to zero as the code length becomes large,
the lower bound asymptotically coincides with the upper one.

When we take $C \setminus \{ \bm{0} \}$ as a trial set $T$, the condition for a lower bound can be weaker and
the lower bound can be improved.
\medskip
\begin{theorem}
Let $C$ be a linear code with even minimum distance $d$.
If 
\begin{equation*}
\frac{1}{2}\binom{d}{\frac{d}{2}} > \left\lceil \frac{|A_d(C)|-1}{2} \right\rceil
\end{equation*} 
holds, then
\begin{multline*}
\frac{1}{2}\binom{d}{\frac{d}{2}}|A_d(C)| - \left\lceil \frac{|A_d(C)|-1}{2} \right\rceil |A_{d}(C)|
 \leq |E^1_{\frac{d}{2}}(C)|.
\end{multline*} 
\end{theorem}
\medskip
\begin{proof}
From Lemma~\ref{lem:evencommon}, a codeword $\bm{c} \in A_d(C)$ has at most one common larger half for 
$\bm{c}' \in A_{d}(C) \setminus \{ \bm{c} \}$.
If $\bm{c}$ and $\bm{c}'$ have the common larger half $\bm{v} \in LH^-(\bm{c}) \cap LH^-(\bm{c}')$, then $\bm{v}$ is
represented as $\bm{v} = \bm{c} \cap \bm{c}'$ and it holds that $l(\bm{c}) = l(\bm{c}')$.
Then the other codeword $\bm{c}+\bm{c}' \in A_d(C)$ does not have common larger halves with $\bm{c}$,
since $l(\bm{c}+\bm{c}') \neq l(\bm{c})$.
Therefore at least $|LH^-(\bm{c})| - \lceil(|A_{d}(C)|-1)/2\rceil$ vectors in $LH^-(\bm{c})$ does not have common larger halves.
Thus we have the lower bound $(\binom{d}{d/2}/2-\lceil (|A_d(C)|-1)/2\rceil)|A_d(C)|$.
\end{proof}
\medskip

In what follows, we see that some BCH codes, Reed-Muller codes, and random linear codes satisfy 
the conditions~(\ref{eq:oddcond}) or~(\ref{eq:evencond}).
For an $(n, k)$ linear code $C$, which has code length $n$ and dimension $k$, we choose $C \setminus \{ \bm{0} \}$ as a trial set for $C$.

\subsubsection{Primitive BCH codes}

By using the weight distribution~\cite{desaki97},
we can verify that the $(n, k)$ primitive BCH codes satisfy the condition~(\ref{eq:oddcond})
for $n =127, k \leq 64$ and $n = 63, k \leq 24$.

\subsubsection{Extended Primitive BCH codes}

By using the weight distribution~\cite{desaki97},
we can verify that the $(n, k)$ extended primitive BCH codes satisfy the condition~(\ref{eq:evencond})
for $n =128, k \leq 64$ and $n = 64, k \leq 24$.

\subsubsection{Reed-Muller codes}

For the $r$-th order Reed-Muller code of length $2^m$, the minimum distance is $2^{m-r}$ and 
the number of minimum weight codewords $|A_{2^{m-r}}({\rm RM}_{m,r})|$ is presented in Theorem~9 of~\cite[Chapter 13]{macwilliams},
which is upper bounded by $(2^{m+1}-2)^r$.
Then, for a fixed $r$, the condition~(\ref{eq:evencond}) is satisfied except for small $m$.
Table~\ref{tb:rmtrial} shows which parameters meet the condition~(\ref{eq:evencond}).

\begin{table}[t]
\caption{The $r$-th order Reed-Muller code of length $2^m$ satisfying~(\ref{eq:evencond}).}
\label{tb:rmtrial}
\begin{center}
\begin{tabular}{ll}\hline
$r$ & \multicolumn{1}{c}{$m$} \\\hline
1 & $\geq 4$\\
2 & $\geq 6$\\
3 & $\geq 8$\\
4 & $\geq 10$\\
5 & $\geq 11$\\
6 & $\geq 13$\\\hline
\end{tabular}
\end{center}
\end{table}

The fraction $|A_{d}(C)|/\binom{d}{d/2}$ is upper bounded by
\begin{equation*}
\frac{|A_{d}(C)|}{\binom{d}{d/2}}  \leq \frac{(2^{m+1}-2)^r}{2^{2^{m-r}}} \leq 2^{(m+1)r - 2^{m-r}}.
\end{equation*}
Thus for a fixed $r$ the fraction tends to zero as $m$ becomes large.
This means the upper and lower bounds in Theorem~\ref{th:evenlower} asymptotically coincide.

\subsubsection{Random Linear Codes}

A random linear code is a code whose generator matrix has equiprobable entries.
That is, first we set a parameter $(n, k)$, and then we choose a generator matrix from all the $2^{nk}$ possible generator matrices
with probability $2^{-nk}$.
It is known that with high probability the minimum distance equals to $n\delta_{\rm GV}$, where $1-H(\delta_{\rm GV})=k/n$ and
$H(x)$ is the binary entropy function of $x$~\cite{gilbert52, varshamov57}.
Also it is known that the weight distribution equals the binomial distribution.
Then, $|A_d(C)| \approx (2^k-1)\binom{n}{d}2^{-n}
 \approx \binom{n}{n\delta_{\rm GV}}2^{k-n}
 \approx 2^{n(H(\delta_{\rm GV})+k/n-1)} \approx 1,$
where we use the approximation $\binom{n}{n\lambda} \approx 2^{H(\lambda)}$, and
$|A_{d+1}(C)| \approx (2^k-1)\binom{n}{d+1}2^{-n}
 \approx \binom{n}{n\delta_{\rm GV}}2^{k-n}(n-d)/(d+1)
 \approx 2^{n(H(\delta_{\rm GV})+k/n-1)} \approx 1.
$
Since $\binom{d}{d/2}\approx \sqrt{2/\pi d}2^d \approx 2^{n\delta}$ for even $d$ and
$\binom{d}{(d+1)/2}  \approx 1/\sqrt{2\pi(d+1)}2^{d+1} \approx 2^{n\delta}$ for odd $d$,
where $d = n\delta$,
the conditions~(\ref{eq:oddcond}) and~(\ref{eq:evencond}) are satisfied.
Since the fractions $|A_{d+1}(C)|/\binom{d}{(d+1)/2}$ and $|A_{d}(C)|/\binom{d}{d/2}$ tend to zero, 
the upper and lower bounds in Theorems~\ref{th:oddlower} and~\ref{th:evenlower} asymptotically coincide.

\subsection*{Remarks}

Note that the condition~(\ref{eq:evencond}) for $T = C \setminus \{ \bm{0} \}$ is a sufficient condition
under which every codewords with weight $d$ is contained in every trial set for $C$ with even minimum distance $d$.
Also,  the condition~(\ref{eq:oddcond}) for $T = C \setminus \{ \bm{0} \}$ is a sufficient condition
under which every codewords with weights $d$ and $d+1$ is contained in every trial set for $C$ with odd minimum distance $d$.

When the condition~(\ref{eq:evencond}) holds for $T = C \setminus \{ \bm{0} \}$,
as described in the proof of Theorem~\ref{th:evenlower},
for every $\bm{c} \in A_d(C)$ there exists at least one larger half $\bm{v} \in LH_{\lceil d/2 \rceil}(T)$
that has no common larger half with other codewords in $C$.
Since $M^1_{\lceil d/2 \rceil}(C) = LH_{\lceil d/2 \rceil}(T)$, every larger half of $\bm{c} \in A_d(C)$ is a minimal uncorrectable error.
Every trial set $T$ must satisfy that $M^1(C) \subseteq LH(T)$.
Therefore, every codeword in $A_d(C)$ needs to be contained in every trial set for $C$ in this case.
By a similar argument, we can show that if the condition~(\ref{eq:oddcond}) for $T = C \setminus \{ \bm{0} \}$ holds,
then every codeword in $A_d(C) \cup A_{d+1}(C)$ need to be in every trial set for $C$.

\section{A Generalization of the Bound}\label{sec:generalize}

By generalizing the results in the previous section, 
we give a lower bound on the size of $LH_i(C \setminus \{ \bm{0} \})$ for each $i$.
We have the relation $M_i^1(C) \subseteq LH_i(C \setminus \{ \bm{0} \}) \subseteq E_i^1(C)$.
Thus the following lower bound is also a lower bound on the number of uncorrectable errors,
but the bound is weak when $i$ is large.
\medskip
\begin{theorem}\label{th:lhlower}
Let $C$ be a linear code with minimum distance $d$ and $T$ be a trial set for $C$.
Define $B_i = |A_{2i-2}(T)| + |A_{2i-1}(T)| + |A_{2i}(T)|$.
For an integer $i$ with $\lceil d/2 \rceil \leq i \leq \lfloor n/2 \rfloor$, if
\begin{equation*}
\binom{2i-3}{i} > 3 \binom{2i- \lceil \frac{d}{2} \rceil}{i} B_i
\end{equation*} 
holds, then
\begin{multline*}
\left(\binom{2i-3}{i} - 3 \binom{2i- \lceil \frac{d}{2} \rceil}{i}B_i\right)B_i
\leq LH_i(T)\\ 
\leq \binom{2i-3}{i}|A_{2i-2}(T)| + 2\binom{2i-1}{i}(|A_{2i-1}(T)| + |A_{2i}(T)|)
\end{multline*} 
\end{theorem}
\medskip
\begin{proof}
First we observe that $LH_i(T) = LH^+(A_{2i-2}(T)) \cup LH(A_{2i-1}(T)) \cup LH^-(A_{2i}(T))$.
We consider the upper bound on the number of common larger halves in $LH_i(T)$.
Let $\bm{c}, \bm{c}'$ be codewords in $A_{2i-2}(T) \cup A_{2i-1}(T) \cup A_{2i}(T)$.
Then $w(\bm{c} \cap \bm{c}') = (w(\bm{c})+w(\bm{c}')-w(\bm{c}+\bm{c}'))/2 \leq (2i+2i-d)/2 = 2i-\lceil d/2 \rceil$.
Therefore the number of common larger halves of weight $i$ between $\bm{c}$ and $\bm{c}'$ is at most $\binom{2i- \lceil d/2 \rceil}{i}$.

For $\bm{c} \in A_{2i-2}(T) \cup A_{2i-1}(T) \cup A_{2i}(T)$, the size of larger halves of $\bm{c}$ with weight $i$ is
at least $\binom{2i-3}{i}$.
Since $\binom{2i-3}{i} > 3 \binom{2i- \lceil d/2 \rceil}{i}B_i$, there is at least 
$\binom{2i-3}{i} - 3 \binom{2i- \lceil d/2 \rceil}{i}B_i$ larger halves of $\bm{c}$ with weight $i$ that have no common larger halves.
Thus the lower bound follows.

The upper bound is obtained from the inequality $|LH_i(T)| 
\leq |LH^+(A_{2i-2}(T))| + |LH(A_{2i-1}(T))| + |LH^-(A_{2i}(T))|
\leq \binom{2i-3}{i}|A_{2i-2}(T)| + \binom{2i-1}{i}|A_{2i-1}(T)| + \binom{2i-1}{i}|A_{2i}(T)|$.
\end{proof}

\section{Concluding Remarks}

A lower bound on the number of uncorrectable errors of weight half the minimum distance have been derived 
for binary linear codes.
The conditions for the bound are not too restrictive, some codes including Reed-Muller codes and random linear codes satisfy the conditions.
A key observation for the results is that an uncorrectable error of weight half the minimum distance
is a larger half of some minimum weight codeword.
The lower bound has been generalized to a lower bound on the size of larger halves of a trial set,
but this bound is not a good lower bound on the number of uncorrectable errors for large weight.
Finding a good lower bound on the number of
uncorrectable error is an interesting future work.


\begin{thebibliography}{9}

\bibitem{ashikhmin98} A.~Ashikhmin and A.~Barg, ``Minimal vectors in linear 
codes,'' {\em IEEE Trans. Inform. Theory}, vol.~44, no.~5, pp.~2010--2017,
Sept.~1998.

\bibitem{desaki97} Y.~Desaki, T.~Fujiwara, and T.~Kasami, ``The weight distributions of extended
binary primitive BCH codes of length 128,'' {\em IEEE Trans. Inform. Theory}, vol.~43, no.~4, July, 1997.

\bibitem{gilbert52} E.N.~Gilbert, ``A comparison of signalling alphabets,''
  {\em Bell System Technical Journal}, vol.~31, pp.~504--522, 1952.

\bibitem{helleseth97} T.~Helleseth and T.~Kl\o ve, ``The Newton radius of codes,''
{\em IEEE Trans. Inform. Theory}, vol.~43, no.~6, pp.~1820--1831, Nov. 1997.

\bibitem{helleseth05} T.~Helleseth, T.~Kl\o ve, and V.~Levenshtein, 
``Error-correction capability of binary linear codes,''
{\em IEEE Trans. Inform. Theory}, vol.~51, no.~4, pp.~1408--1423, Apr. 2005.

\bibitem{macwilliams} F.J.~MacWilliams and N.J.A.~Sloane, 
{\em The theory of error-correcting codes}, North-Holland, 1977.

\bibitem{peterson} W.W.~Peterson and E.J.~Weldon,~Jr., 
{\em Error-Correcting Codes, 2nd Edition}, MIT Press, 1972.

\bibitem{poltyrev94} G.~Poltyrev, ``Bounds on the decoding error probability of binary linear codes via their spectra,''
{\em IEEE Trans. Inform. Theory}, vol.~40, no.~4, pp.~1284--1292, July 1994.

\bibitem{varshamov57} R.R.~Varshamov, ``Estimate of the number of signals in
  error correcting codes,'' {\em Doklady Akadamii Nauk SSSR}, vol.~117, pp.~739--741, 1957.

\bibitem{wu98} C.K.~Wu,``On distribution of Boolean functions with nonlinearity $\leq 2^{n-2}$'',
{\em Australasian Journal of Combinatorics}, vol.~17, pp.~51--59, Mar. 1998.

\bibitem{yasunaga07} K.~Yasunaga and T.~Fujiwara, ``Correctable errors of weight half the minimum distance plus one 
for the first-order Reed-Muller codes,'' in {\em Proc. Applied Algebra, Algebraic Algorithms, and Error Correcting Codes, 
Lecture Notes in Computer Science}, vol.~4851, Springer, pp.~110--119, Dec. 2007.

\bibitem{zemor93} G.~Z\'emor, ``Threshold effects in codes,''
in {\em Proc. Algebraic Coding}, Paris, France, 1993,
{\em Lecture Notes in Computer Science}, vol.~781, Springer, pp.~278--286, 1994.







\end{thebibliography}
\end{document}